# Hall effect in the accumulation layers on the surface of organic semiconductors

V. Podzorov,[1,*] E. Menard,[2,3] J. A. Rogers,[2] and M. E. Gershenson[1]

[1] *Department of Physics and Astronomy, Rutgers University, Piscataway, New Jersey, USA*
[2] *Department of Materials Science and Engineering, University of Illinois, Urbana-Champaign, Illinois, USA*
[3] *CEA-Saclay, LCR Semi-Conducteurs organiques, SPCSI/DRECAM/DSM, F-91191 Gif-sur-Yvette (France)*

(July 28, 2005)

We have observed the Hall effect in the field-induced accumulation layer on the surface of small-molecule organic semiconductor. The Hall mobility $\mu_H$ increases with decreasing temperature in both the intrinsic (high-temperature) and trap-dominated (low-temperature) conduction regimes. In the intrinsic regime, the density of mobile field-induced charge carriers extracted from the Hall measurements, $n_H$, coincides with the density $n$ calculated using the gate-channel capacitance, and becomes smaller than $n$ in the trap-dominated regime. The Hall data are consistent with the diffusive band-like motion of field-induced charge carriers between the trapping events.

*Electronic mail: podzorov@physics.rutgers.edu

After several decades of intensive research, our understanding of the charge transport in small-molecule organic semiconductors remains incomplete. Complexity of the transport phenomena in these systems is due to the polaronic nature of charge carriers [1] and the strong interaction of small polarons with defects. For the emerging field of organic electronics [2], it is especially important to develop an adequate model of the polaronic transport at *room temperature*. This is a challenging task because the energy of thermal excitations at room temperature may be comparable to the width of the conduction band in these van-der-Waals-bonded materials. The recent theories [3,4,5,6,7] show that the high-$T$ polaronic transport in organic semiconductors is governed by the competition between the bandwidth narrowing that would lead to a decrease of the carrier mobility $\mu$ with $T$ and the thermally-activated hopping processes that result in an increase of $\mu$ with $T$. Accordingly, the crossover from the band-like transport in delocalized states to the inelastically-assisted incoherent hopping between localized states is expected with increasing temperature.

Recent development of *single-crystal* organic transistors (OFETs) [see, e.g., 8] with significantly reduced disorder enabled realization of the intrinsic (not limited by static disorder) polaronic transport on organic surface [9]. The room-temperature mobility of carriers in the rubrene-based single-crystal OFETs (up to 20 cm$^2$/Vs) exceeds tenfold the RT mobility measured in the time-of-flight experiments with the bulk crystals of naphthalene and anthracene [10], the benchmark results for the intrinsic transport in organic semiconductors. However, even this high (for organic semiconductors) value of $\mu$ does not guarantee that the mean free path, $l$, significantly exceeds the intermolecular distance, the necessary condition for the diffusive band-like transport (see below).

The Hall measurements may shed light on this complicated problem. Indeed, the magnitude and temperature dependence of the Hall effect are expected to be qualitatively different for the diffusive transport in delocalized states and for the incoherent thermally-activated hopping [11,12]. In the diffusive regime, the Hall constant $R_\mathrm{H} = 1/(en)$ is inversely proportional to the two-dimensional density of mobile carriers in the accumulation layer, $n$. For hopping between localized states, it is not possible to introduce a classical velocity of carriers and, thus, the Lorentz force. The Hall effect in the hopping regime may arise from a quantum interference mechanism [12]; it is expected that the magnitude of the Hall voltage, strongly suppressed in comparison with the diffusive regime, acquires an exponentially-strong temperature dependence. In the scarce measurements of the Hall effect in the hopping regime for conventional semiconductors (see, e.g., [13,14]), very small Hall voltages and the sign anomaly of the Hall constant have been observed.

In this Letter, we report on observation of the Hall effect in the electric-field-induced accumulation layers on the surface of small-molecule organic semiconductor. For these experiments, we have used the organic field-effect transistors (OFETs) based on single crystals of *rubrene*. The Hall data were obtained over a wide temperature range that spans over the intrinsic (high-$T$) and trap-dominated (low-$T$) conduction regimes [9]. At high temperatures, where trapping by shallow traps is negligible, the carrier density $n_H$ extracted from the Hall measurements coincides with the density of the field-induced carriers $n$ estimated from the gate-channel capacitance. The Hall mobility increases with decreasing temperature in both intrinsic and trap-dominated regimes. Our data suggest that the charge transport on the surface of rubrene single crystals occurs via delocalized states over the whole studied temperature range up to room temperature.

To study the Hall effect, we have used two types of the field-effect transistors based on vapor-grown organic molecular crystals: the devices with the polymer parylene film as a gate dielectric [15] and the elastomeric stamp-based devices with the micron-size gap between the surface of organic crystal and the gate electrode [9] (referred below as the "vacuum-gap" OFETs). Typical dimensions of the structures are (see the inset in Fig. 1): the channel length $L$ = 1-3 mm, channel width $W = 0.2 – 1.4$ mm, the separation between the voltage contacts 1 and 2 in the 4-probe conductivity measurements $L^* = 0.3 – 0.6$ mm. The gate-channel capacitance per unit area, $C_i$, is ~ 2.1 nF/cm$^2$ for the devices with the parylene dielectric and ~ 0.2 nF/cm$^2$ for the vacuum-gap OFETs. The OFET conduction channel was oriented along the ***b***-axis of the crystal. We have used Keithley source-meters K2400 and electrometers K6512 for the measurements of the 4-probe voltage $V$ between the contacts 1 and 2, the Hall voltage $V_H$, and the source-drain current $I$. Qualitatively the same data have been obtained for both types of devices; below we present the measurements for one of the vacuum-gap OFETs. The measurements were conducted over the temperature range $T = 150 – 300$ K in magnetic fields $B = 0 - 6$ T.

Figure 1 shows how the voltage $V_H$, measured between the Hall contacts at fixed gate ($V_G$) and source-drain ($V_{SD}$) voltages, varies with time ($t$) when magnetic field $B(t)$ is applied perpendicular to the channel. An offset voltage originated from a small asymmetry of the Hall probes exhibits a slow monotonic drift. This offset voltage does not depend on $B$: interestingly, we were unable to detect any longitudinal magnetoresistance $\Delta\sigma(B)$ within the accuracy of our measurements. Because of the slow monotonic drift of the offset voltage, the Hall voltage $U_H$ was determined by subtracting the $B$-independent offset from $V_H(t)$: $U_H(B) \equiv [V_H(B) - V_H(-B)]/2$. The signal-to-noise ratio and, therefore, the accuracy of $U_H$ measurements increases with $|V_G|$. Figure 2 shows that the Hall voltage $U_H$ is proportional to the applied magnetic field and changes its sign when the direction of $B$ is reversed. Its magnitude is almost $V_G$-independent and increases linearly with the source-drain voltage $V_{SD}$ (the upper and lower insets in Fig. 2, respectively). The sign of $U_H$ is consistent with the p-type conductivity in the studied rubrene OFETs.

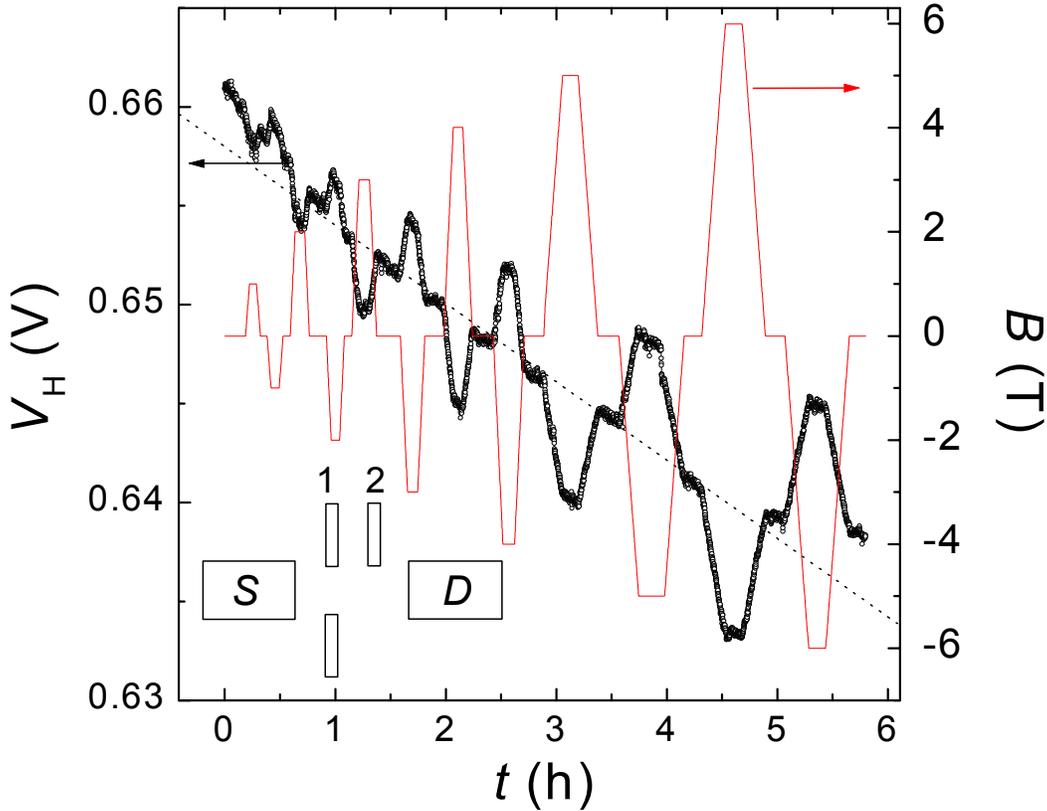

**Figure 1.** Voltage between the Hall contacts $V_H$ (dots) recorded as a function of time for a rubrene single-crystal OFET at fixed $V_{SD} = 5$ V, $V_G = -40$ V, and $T = 300$ K. The time dependence of the external magnetic field is shown by the solid line. The inset shows the contact geometry.

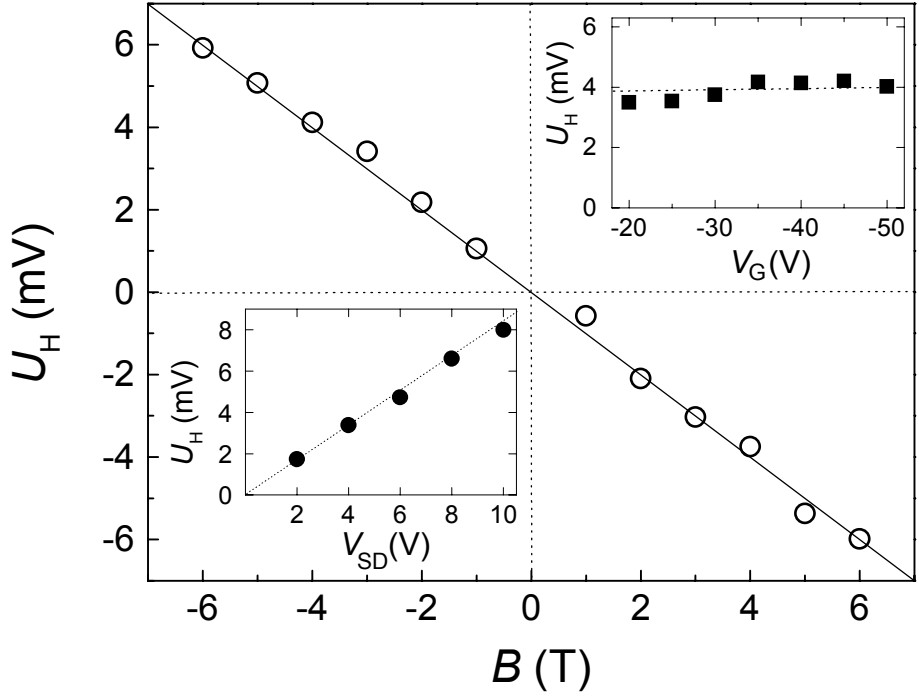

**Figure 2**. The dependence of the Hall voltage $U_H$ on the magnetic field $B$ ($T$ = 300 K, $V_{SD}$ = 5V and $V_G$ = -40 V). The insets show the dependences of $U_H$ on the gate voltage, $V_G$ and on the source-drain voltage, $V_{SD}$.

From the dependences $U_H(B)$, we calculated the Hall constant, $R_H$, the Hall carrier density, $n_H$, and the Hall mobility, $\mu_H$ [11]:

$$R_H \equiv \frac{1}{B}\frac{U_H}{I} \quad , \quad (1)$$

$$n_H \equiv \frac{1}{eR_H} = B\frac{I}{eU_H} \quad , \quad (2)$$

$$\mu_H \equiv R_H \sigma = \frac{1}{B}\frac{U_H}{V}\left(\frac{L^*}{W}\right) \quad . \quad (3)$$

Here σ is the channel conductivity (measured by the 4-probe technique to account for the contact resistance) and $V$ is the voltage between the voltage probes 1 and 2 in the 4-probe geometry. The temperature dependence of $\mu_H$ is shown in the upper panel of Fig. 3, along with the effective mobility $\mu_{eff}$ extracted from the 4-probe FET measurements of the conductivity (see below). We emphasize that $\mu_H$ does not depend on $V_G$ (i.e., on the density of charge carriers in the channel) over the whole experimental $T$ range; the value of $\mu_H$ at room temperature (~ 10 cm$^2$/Vs) is consistent with our previous measurements of the mobility in rubrene OFETs along the ***b*** axis [9]. The lower panel of Fig. 3 shows the temperature dependence of the Hall density $n_H$, normalized by the density of charge carriers $n$ field-induced in the channel above the threshold voltage ($|V_G| > |V_G^{th}|$):

$$n = \frac{C_i}{e}\left[V_G - V_G^{th}(T)\right]. \qquad (4)$$

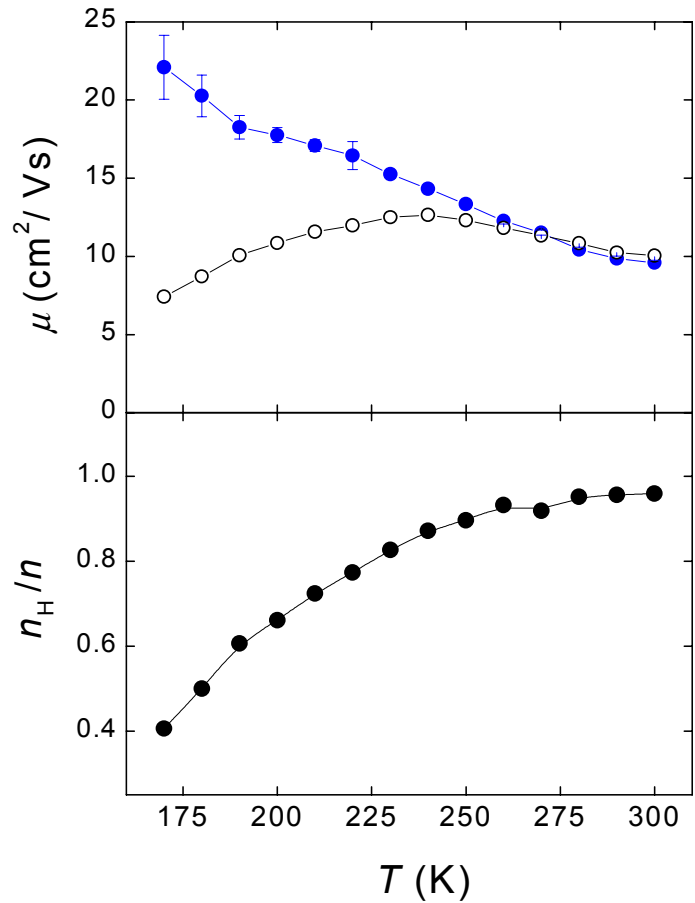

**Figure 3**. Upper panel: the temperature dependence of the Hall mobility, $\mu_H$, (solid circles) and the mobility extracted from the conductivity σ using the density $n$ calculated from the gate-channel capacitance (open circles). Lower panel: the temperature dependence of the ratio of the Hall carrier density, $n_H$, to the density $n$.

At a fixed $V_G$, $n$ decreases with cooling due to a quasi-linear increase $V_G^{th}(T)$ which is proportional to the density of deep traps [9]: e.g., at $V_G = -40$ V, $n$ decreases from $4.5 \cdot 10^{10}$ cm$^{-2}$ at $T = 300$ K to $2 \cdot 10^{10}$ cm$^{-2}$ at $T = 170$ K. The systematic error of determining the values of $n_H$ and $n$ does not exceed ~ 15%.

The Hall measurements in the studied devices are limited at lower $T$ by a rapid growth of the fluctuations of $V_H$ (note the error bars in the upper panel of Fig. 3). These fluctuations are related to the noise of the "background" offset voltage, which does not depend on the applied magnetic field (Fig. 4). We found that the power density of this noise $S_V \equiv \langle (V - \langle V \rangle)^2 \rangle$ exhibits the 1/$f$ frequency dependence. The $T$ dependence of the normalized density of the 1/$f$ noise, $S_V/V^2$, which presumably reflects the fluctuations of the channel resistance, is shown in Fig. 4. In the intrinsic conduction regime ($T \geq 240$ K), the noise density $S_V/V^2$ is $T$-independent, whereas in the trap-dominated regime, it increases dramatically with decreasing temperature. The magnitude of 1/$f$ noise decreases with the applied negative gate voltage and, thus, with the increase of charge density in the channel. The nature of these fluctuations requires further studies; at this stage, we can only speculate that the fluctuations may be related to the trapping-related fluctuations of the number of mobile charge carriers in the channel.

Let us start the analysis of the experimental data with the discussion of the OFET conductivity. The charge carriers field-induced above the threshold ($|V_G| > |V_G^{th}|$) participate in the current flow along the conduction channel. According to the multiple trap-and-release (MTR) model (see, e.g., [16]), these carriers can be trapped over the time scale $\tau_{tr}$ by the shallow traps (i.e., the trap states with the energies within a few $k_B T$ above the HOMO band). The trapping time increases exponentially with decreasing $T$: $\tau_{tr} \propto \exp(E_a/T)$, where $E_a \sim 70$ meV is a characteristic energy scale reflecting the energy distribution of tail states for the studied devices [9]. The effect of trapping on the channel conductivity $\sigma = en\mu$ can be described using two approaches. According to a more conventional approach (see, e.g., [9]), all the carriers at a density $n$ contribute to the current flow, but, because of trapping, the effective mobility of these carriers is reduced in comparison with its "intrinsic", trap-free value $\mu_0$:

$$\mu_{eff} \equiv \frac{\sigma}{en} = \mu_0 \frac{\tau}{\tau + \tau_{tr}}. \quad (5)$$

Here $\tau$ is the average time that a polaron spends traveling between shallow traps. For the diffusive motion, $\tau \sim d^2/D$ ($d$ is the distance between the shallow traps, $D$ is the diffusion constant) is inversely proportional to the density of shallow traps $N$. In the studied vacuum-gap OFETs, $N \sim 10^{10}$ cm$^{-2}$ and $d \sim 10^{-5}$ cm [9]. According to Eq. 5, the intrinsic regime of conduction realizes when $\tau \gg \tau_{tr}$; in the opposite limit, the transport is dominated by trapping events. Alternatively, in the second approach, one can take into account that among $n$ charges field-induced above the threshold (Eq. 4), only a certain number is mobile, namely

$$n_{eff} = n \frac{\tau}{\tau + \tau_{tr}} \qquad (6),$$

and the motion of these charges is characterized with the intrinsic mobility $\mu_0$. The other charges with the concentration $n - n_{eff}$ are temporarily immobilized by the shallow traps.

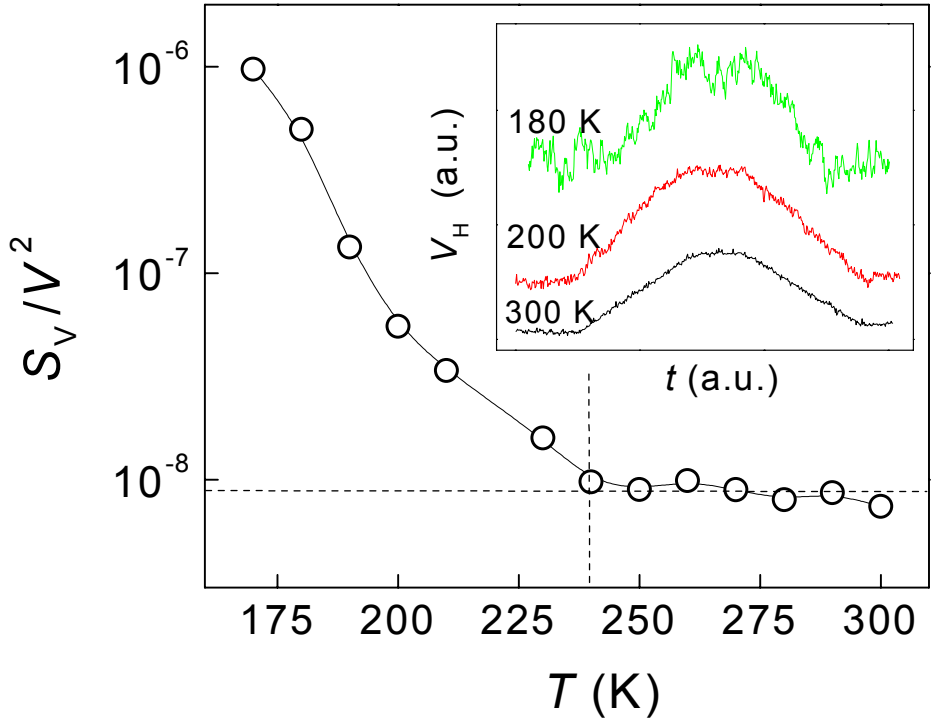

**Figure 4**. The temperature dependence of the normalized power density of $1/f$ noise of the "background" offset voltage caused by the asymmetry of Hall probes. The inset shows $V_H$ recorded as a function of time at several temperatures, when $B$ was swept from 0 to 6 T and back to 0.

Both approaches are equivalent when one analyzes $\sigma = en\mu$. An advantage of the Hall measurement is that it allows *independent* measurements of $n$ and $\mu$. The quantity $n$ that is determined in the Hall measurements is the density of charges that are moving at a given moment of time, i.e., $n_H = n_{eff}$. Indeed, the charges that are temporarily trapped in shallow traps do not contribute to the Hall voltage. The mobility extracted from the Hall experiments should coincide with the intrinsic, trap-free mobility $\mu_0$. These expectations are in line with the observed $T$-dependences of $n_H$ and $\mu_H$. The values of $n_H$ and $n$ (Eqs. 2 and 4) coincide with each other within the experimental accuracy at $T > 240$ K, where the effect of trapping is negligible (the intrinsic regime). In the trap-dominated regime ($T < 240$ K), $n_H$ becomes smaller than $n$ as $\tau_{tr}$ increases and exceeds $\tau$ (see the lower panel in Fig.3). At the same time, the intrinsic mobility $\mu_0$ determined in the Hall measurements continues to increase with cooling even at low $T$, where the effective mobility $\mu_{eff}$, being significantly affected by trapping, decreases with cooling.

The observed agreement between the room-temperature values of $n$ extracted from the Hall constant (Eq. 2) and calculated from the gate-channel capacitance (Eq. 4) is characteristic for the *band-like* transport in delocalized states. As far as we know, the observation of non-activated transport in a two–dimensional system with the RT sheet resistance $R \sim 12$ MΩ (at $V_G = -40$V) is unique: typically, the charge transport at $R \gg h/e^2 = 25.8$ kΩ is governed (at least at low temperatures) by the thermally-activated hopping between localized states. The diffusive nature of high-$T$ polaronic transport in rubrene is in line with the predictions made on the basis of band calculations for rubrene [6] and the *ab initio* calculations of the mobility of small polarons [7]. According to the latter theory, the diffusion should dominate in the p-type conductivity in the small-molecular crystals up to the room temperature, whereas the contribution of hopping remains small.

It is commonly believed that for the realization of diffusion motion in delocalized states, the mean free path of carriers, $l = m^*\mu v/e$ ($v$ is the carrier velocity and $m^*$ is their effective mass), should exceed the intermolecular distance (~ 1 nm in the case of rubrene). Taking into account that the statistics of polarons is non-degenerate at $n \sim 5\cdot10^{-10}$ cm$^{-2}$ and $T = 300$K, and that they move with the thermal velocity $v = (3k_BT/m^*)^{1/2}$, one can obtain a lower limit of the effective mass of polaronic carriers, $m^*$:

$$m^* = \left(\frac{el}{\mu}\right)^2 \frac{1}{3k_B T} \geq 2m_e \qquad (7)$$

where $m_e$ is the bare electron mass.

To summarize, we have observed the Hall effect in the field-effect structures based on single crystals of a small-molecule organic semiconductor rubrene. The Hall experiment enabled the first direct measurement of the density of mobile carriers in the conduction channel and their intrinsic trap-free mobility over a wide temperature range. In the intrinsic regime, the density of mobile charges is in a good agreement with the density calculated from the gate-channel capacitance. The Hall measurements provide the data on the intrinsic mobility of polarons even at low temperatures, where the charge transport is dominated by trapping. These findings suggest that the model of diffusive (non-activated) transport in the delocalized states is applicable to the motion of mobile polaronic charges on the surface of organic semiconductors up to the temperatures as high as 300 K.

This work has been supported by the NSF grants DMR-0405208 and ECS-0437932.

1. E. A. Silinsh and V. Čápek, *Organic Molecular Crystals: Interaction, Localization, and Transport Phenomena* (AIP Press, New York, 1994);
2. S. R. Forrest, *Nature*, **428**, 911 (2004).
3. M. W. Wu and E. M. Conwell, *Chem. Phys. Lett.*, **266**, 363 (1997).
4. W.-Q. Deng and W. A. Goddard, *J. Phys. Chem*. B, **108**, 8614 (2004).
5. V. M. Kenkre *et al.*, *Phys. Rev. Lett*. **62**, 1165 (1989).
6. D. A. da Silva Filho, E.-G. Kim, and J.-L. Bredas, *Adv. Materials* **17**, 1072 (2005).
7. K. Hannewald and P. A. Bobbert, *Appl. Phys. Lett*. **85**, 1535 (2004).
8. R. W. I. de Boer, M. E. Gershenson, A. F. Morpurgo, V. Podzorov, "Organic single-crystal field-effect transistors", *Phys. Stat. Solidi*, **201**, 1302 (2004); J. Takeya *et al.*, *J. Appl. Phys.* **94**, 5800 (2003); C. R. Newman, R. J. Chesterfield, J. A. Merlo, and C. D. Frisbie, *Appl. Phys. Lett*. **85**, 422 (2004).
9. V. Podzorov *et al.*, *Phys. Rev. Lett*. **93**, 086602 (2004); E. Menard *et al.*, *Adv. Mater*. **16**, 2097 (2004).
10. N. Karl in *Organic Electronic Materials*, eds. R. Farchioni and G. Grosso, pp. 215-239 and 283-326 (Springer-Verlag Berlin Heidelberg, 2001).
11. M. Pope and C. E. Swenberg, *Electronic Processes in Organic Crystals and Polymers*. 2nd ed., Oxford University Press, New York, London, pp. 374-378 (1999).
12. L. Friedman and M. Pollak, *Phil. Mag*. **38**, 173 (1978); L. Friedman, *Phil. Mag*. **38**, 467 (1978).
13. D. C. Look *et al.*, *Phys. Rev*. B **42**, 3578 (1990).

14. C. E. Nebel et al., *Phil. Mag. Lett.* **74**, 455 (1996).

15. V. Podzorov, V. M. Pudalov, and M. E. Gershenson, *Appl. Phys. Lett.* **82**, 1739 (2003); V. Podzorov *et al.*, *Appl. Phys. Lett.* **83**, 3504 (2003);

16. G. Horowitz, *J. Mater. Res.*, **19**, 1946 (2004).